# CLAUSIUS' ENTROPY REVISITED

E. G. D. Cohen[1,2,a] and R. L. Merlino[2,b]

[1]*The Rockefeller University, 1230 York Avenue, New York, NY 10065*
[2] *Department of Physics and Astronomy, The University of Iowa, Iowa City, IA 52245*

(January 10, 2014)

## ABSTRACT

Conventional Non-equilibrium Thermodynamics is mainly concerned with systems in local equilibrium and their entropy production, due to the irreversible processes which take place in these systems. In this paper fluids will be considered in a state of local equilibrium. We argue that the main feature of such systems is not the entropy production, but the organization of the flowing currents in such systems. These currents do not only have dissipation or entropy production, but must also have an order or organization needed to flow in a certain direction. It is the latter, which is the source of the equilibrium entropy, when the fluid goes from a local equilibrium state and to an equilibrium state. This implies a transmutation of the local equilibrium currents' organization into the equilibrium entropy. Alternatively, when a fluid goes from an equilibrium state to a local equilibrium state, its entropy transmutes into the organization of the currents of that state.

**Keywords:** non-equilibrium thermodynamics, entropy
**PACS Codes:** 05.70.-a, 05.70.Ln

_______________________
[a] egdc@rockefeller.edu
[b] robert-merlino@uiowa.edu

## I. INTRODUCTION

In this paper we will consider fluids (i.e., gases or liquids) near equilibrium, which are in a non-equilibrium state of local equilibrium, and *vice versa*. In such a fluid, local thermodynamic properties can be defined. This, because it is assumed then that the system can be described in first approximation by a non-equilibrium local Maxwell (LM) velocity distribution function [1]:

$$f_{\mathrm{LM}}\left(\mathbf{r},\mathbf{v},t\right) = c_{\mathrm{LM}}\left(\mathbf{r},t\right) n\left(\mathbf{r},t\right) e^{\frac{-m\left[\mathbf{v}-\mathbf{u}(\mathbf{r},t)\right]^2}{2kT(\mathbf{r},t)}}. \tag{1}$$

Here $c_{\mathrm{LM}}\left(\mathbf{r},t\right) = \left[m/2\pi kT\left(\mathbf{r},t\right)\right]^{3/2}$ is the normalization factor of the local Maxwell distribution function, m is the mass of a fluid particle, $\mathbf{v}$ its velocity and k is Boltzmann's constant. In Equation (1), $\mathbf{r}$ denotes a position in the system and $t$ the time, $n\left(\mathbf{r},t\right)$ the local number density, $\mathbf{u}(\mathbf{r},t)$ the local velocity and $T\left(\mathbf{r},t\right)$ the local temperature of the fluid, respectively:

$$\left. \begin{aligned} n\left(\mathbf{r},t\right) &= \int f_{\mathrm{LM}}\left(\mathbf{r},\mathbf{v},t\right) d\mathbf{v} \\ \mathbf{u}\left(\mathbf{r},t\right) &= \int f_{\mathrm{LM}}\left(\mathbf{r},\mathbf{v},t\right) \mathbf{v}\, d\mathbf{v} \\ \frac{3}{2}kT\left(\mathbf{r},t\right) &= \int f_{\mathrm{LM}}\left(\mathbf{r},\mathbf{v},t\right)\left(\frac{1}{2}m\mathbf{v}^2\right) d\mathbf{v} \end{aligned} \right\} \tag{2}$$

These local quantities are physically interpreted as (coarse-grained) averages over a "physically infinitesimal" volume element around every position $\mathbf{r}$ in the system, at time $t$, which contains a sufficient number of fluid particles to define physically meaningful average local quantities around any point $\mathbf{r}$ at time $t$ in the fluid [2]. Considering the size of a macroscopic system and the enormous number of particles in it, such local "coarse grained" volume elements should be "definable". In the local equilibrium state the gradients of the thermodynamic quantities are small. For example, in the case of heat conduction, $L\,\nabla T\left(\mathbf{r},t\right)/T\left(\mathbf{r},t\right) << 1$. Here $L$ is a characteristic length: the mean free path in a gas or the size of the particles in a liquid, respectively.

Experimentally, a fluid in a local equilibrium state can be obtained by applying appropriate *external forces* to a fluid in an equilibrium state. A simple example is heat conduction, when an isolated system of interest is acted upon by externally attaching to it two heat reservoirs of different temperatures, which will induce a temperature gradient, leading to a heat current in the system from the hot to the cold reservoir.

Only when these external forces are removed, will the fluid return to equilibrium (*eq*), since then, $n\left(\mathbf{r},t\right) \to n_{eq}, \mathbf{u}\left(\mathbf{r},t\right) \to 0, T\left(\mathbf{r},t\right) \to T_{eq}$ with

$$n_{eq} = \int f_M(\mathbf{v}) d\mathbf{v}$$

$$\mathbf{u}_{eq} = 0 \qquad\qquad\qquad (3)$$

$$\frac{3}{2} kT_{eq} = \int f_M(\mathbf{v}) \left(\frac{1}{2} m\mathrm{v}^2\right) d\mathbf{v}$$

with

$$f_M(\mathbf{v}) = n_{eq} c_M e^{-\frac{m\mathrm{v}^2}{2kT_{eq}}}, \qquad\qquad (4)$$

the equilibrium Maxwell (M) velocity distribution function with, $c_M = \left(m/2\pi kT_{eq}\right)^{3/2}$ and $\mathrm{v} = |\mathbf{v}|$ [3].

In the classical book on Non-Equilibrium Thermodynamics by De Groot and Mazur [4], the entropy production is the main interest. In fact, the entropy production is a necessary but only a marginal side effect of the currents, due to their dissipation while flowing, e.g. by viscous friction. This dissipation of the currents *must* be present, because of the Second Law of Thermodynamics' impossibility of a *perpetuum mobile* of the second kind. The connection of the local equilibrium state with the equilibrium state, except for the disappearance of the entropy production in that state, is not mentioned in [4]. On the contrary, the present paper deals with the connection between a fluid in an equilibrium state and a non-equilibrium state and *vice versa*.

## II. ENTROPY PRODUCTION

As mentioned above, the Second Law of Thermodynamics requires an entropy production of the currents in a local equilibrium state. We will restrict ourselves in this paper to one current only, for more currents see [4].

The entropy production of a current per unit time and unit volume $\sigma$ is given by the product of the current $\mathbf{J}$ and its corresponding gradient $X$ [4]

$$\sigma = \mathbf{J} \bullet \mathbf{X} \qquad\qquad (5)$$

where the space and time dependences have been suppressed, as will be done in the rest of this paper. Since in Non-equilibrium Thermodynamics the current is assumed to be linear in the corresponding gradient:

$$\mathbf{J} = \alpha \mathbf{X}, \qquad\qquad (6)$$

Where $\alpha$ a is space and time independent, i.e., constant, positive transport coefficient, corresponding to $\mathbf{J}$. Inserting Equation (6) into Equation (5) gives

$$\sigma = \alpha X^2, \qquad (7)$$

with $X = |\mathbf{X}|$ and $\sigma$ is non-negative, with a minimum of zero in thermal equilibrium. Thus, e.g. in the case of thermal conduction, where there is a heat current

$$\mathbf{J} = \lambda \mathbf{X} = \lambda \nabla T, \qquad (8)$$

with $\lambda$ the corresponding transport coefficient, the thermal conductivity. Therefore,

$$\sigma = \lambda X^2 = \lambda \left( \nabla T \right)^2 \geq 0. \qquad (9)$$

It is important to note that the current in a fluid in a local equilibrium state is proportional to the gradient cf., Equation (6), while the entropy production is proportional to the square of the gradient cf., Equation (7). Since the gradient is small, as mentioned above, the current's entropy production $\sigma$ is much smaller than the magnitude of the current.

### III.  CURRENT ORGANIZATION

In a fluid in local equilibrium, the velocities of the fluid particles in a current necessarily possess some order or *organization* [5], since they flow and therefore must move, on average, in the same direction [6]. This can be observed, e.g., in the streamlines [7] in a laminar flow of e.g., the Poiseuille flow [8] shown schematically Figure 1.

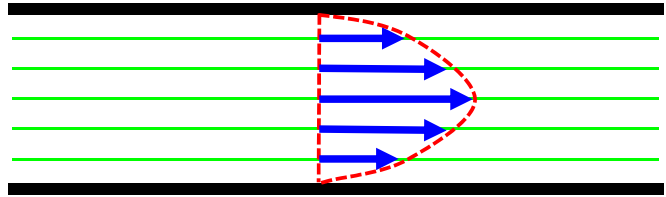

**Figure 1.** Streamlines (green lines) of a laminar Poiseuille flow in a pipe [8]. The blue arrows represent the local fluid velocity.

It is the *current* which is the most important aspect of the local equilibrium state, since it has both organization as well as dissipation. For that reason the local equilibrium state is physically much more complicated than the equilibrium state. This, because the Maxwell velocity distribution function Equation (4) implies just a random Gaussian distribution of the velocities of the fluid particles, which inhibits organization of the fluid [9]. It seems to have been overlooked in Non-equilibrium Thermodynamics, as in [4], that there must be an internal structure associated with the current as incorporated in their organization [10].

## IV. TRANSITION FROM LOCAL EQUILIBRIUM TO EQUILIBRIUM AND *VICE-VERSA*

The organization of a current is crucial for the following reason: when the external forces are removed, and the fluid in a local equilibrium state goes to an equilibrium state, the entropy production vanishes when it has reached equilibrium, because the gradient then disappears, cf., Equation (7). There seems then to be *no* source to account for the (maximum) entropy of the equilibrium state.

The solution of this problem is the above-mentioned organization of the fluid particles in a current. For, while the fluid goes to equilibrium and the gradient gradually disappears, the current will more and more disintegrate and lose its organization, so that the fluid will increasingly randomize. When the equilibrium state is reached, its randomization will be maximal, which provides the maximum equilibrium entropy, as required by Equilibrium Thermodynamics. Thus the local equilibrium current's organization has been *transmuted* into the equilibrium entropy. Alternatively, when external forces are applied to a fluid in equilibrium to bring it into a local equilibrium state, its entropy is transmuted into the organization of the current in the local equilibrium state.

Thus, Clausius' somewhat mysterious and (invisible) entropy, a measure of the particles' disorder in a system in equilibrium, undergoes a *transmutation* into the particles' (visible) organization in a local equilibrium state. One could call therefore the organization of a current: "antropy", from anti-entropy, since their "transmutability" shows their intimate physical connection.

## V. CONNECTION WITH INFORMATION THEORY

The transmutation of the entropy in the equilibrium state to the current organization in the local equilibrium state, can also be seen as a physical example of Landauer's Principle of Information Theory [11] involving a non-equilibrium in addition to an equilibrium system [12]. This principle reads that: "To erase information, energy is needed". Here, the maximum information contained in the equilibrium entropy transmutes into the much lower information in the non-equilibrium currents' organization. Thus, Landauer's principle requires that energy has to be supplied, or work has to be done, to go from an equilibrium state to a non-equilibrium state. Since according to Equilibrium Thermodynamics the energy of a system in equilibrium is a minimum, the transmutation of the equilibrium entropy into the local equilibrium current organization requires work, i.e., energy, in agreement with Landauer's Principle.

## ACKNOWLEDGEMENT

One of us (EGDC) is indebted to Professor Larry Sirovich for helpful discussions on Information Theory.